\documentclass[a4paper,11pt]{article}
\usepackage{pos}

\title{Relentless multi-wavelength variability of Mrk\,421 and Mrk\,501}
\ShortTitle{Relentless multi-wavelength variability of Mrk\,421 and Mrk\,501}

\author*[a]{Vitalii~Sliusar}
\author[b]{Axel~Arbet-Engels}
\author[c]{Dominik~Baack}
\author[a]{Matteo~Balbo}
\author[b]{Adrian~Biland}
\author[b,e]{Thomas~Bretz}
\author[c]{Jens~Buss}
\author[d]{Daniela~Dorner}
\author[d]{Laura~Eisenberger}
\author[c]{Dominik~Elsaesser}
\author[b]{Dorothee~Hildebrand}
\author[d]{Roman~Iotov}
\author[d]{Adelina~Kalenski}
\author[d]{Karl~Mannheim}
\author[b]{Alison~Mitchell}
\author[b]{Dominik~Neise}
\author[c]{Maximilian~Noethe}
\author[d]{Aleksander~Paravac}
\author[c]{Wolfgang~Rhode}
\author[\,d]{Bernd~Schleicher}
\author[]{Roland~Walter$^a$ (the~FACT~Collaboration$^\dagger$)}

\affiliation[a]{Department of Astronomy, University of Geneva,\\ Chemin d'Ecogia 16, CH-1290 Versoix, Switzerland}
\affiliation[b]{Department of Physics, ETH Zurich,\\ Otto Stern Weg 5, CH-8093 Zurich, Switzerland}
\affiliation[c]{Fakult\"at Physik, Technische Universit\"at Dortmund,\\ Otto Hahn Str. 4a, D-44227 Dortmund, Germany}
\affiliation[d]{Institut f\"ur Theoretische Physik und Astrophysik, Universit\"at W\"urzburg,\\ Emil Fischer Str. 31, 97074 W\"urzburg, Germany}
\affiliation[e]{Physikalisches Institut III A, RWTH Aachen University, Otto Blumenthal Str., D-52074 Aachen, Germany

\footnotetext[2]{\href{https://fact-project.org/}{https://fact-project.org/}. For the collaboration list see \href{https://fact-project.org/collaboration/icrc2021_authorlist.html}{https://fact-project.org/collaboration/icrc2021\_authorlist.html}}
}

\emailAdd{vitalii.sliusar@unige.ch}

\abstract{
Mrk\,421 and Mrk\,501 are two close, bright and well-studied high-synchrotron-peaked blazars, which feature bright and persistent GeV and TeV emission. We use the longest and densest dataset of unbiased observations of these two sources, obtained at TeV and GeV energies during five years with FACT and {\it Fermi}-LAT. To characterize the variability and derive constraints on the emission mechanism, we augment the dataset with contemporaneous multi-wavelength observations from radio to X-rays. We correlate the light curves, identify individual flares in TeV energies and X-rays, and look for inter-band connections, which are expected from the shock propagations within the jet. For Mrk\,421, we find that the X-rays and TeV energies are well correlated with close to zero lag, supporting the SSC emission scenario. The timing between the TeV, X-ray flares in Mrk\,421 is consistent with periods expected in the case of Lense–Thirring precession of the accretion disc. The variability of Mrk\,501 on long-term periods is also consistent with SSC, with a sub-day lag between X-rays and TeV energies. Fractional variability for both blazars shows a two bump structure with the highest variability in the X-ray and TeV bands.
}

\FullConference{37$^{\rm{th}}$ International Cosmic Ray Conference (ICRC 2021)\\
		July 12th -- 23rd, 2021\\
		Online -- Berlin, Germany}

\begin{document}

\maketitle

\section{Introduction\label{sec:introduction}}

\vspace*{-0.5em}
Blazars are the most extreme class of active galactic nuclei (AGN). These sources are radio-loud AGN featuring a flat radio spectrum. Their emission is non-thermal spanning from radio to very-high-energy $\gamma$-rays. The radiation is associated with a relativistic jet aligned close to the observer's line of sight. Due to the effects of relativistic beaming, jets in blazars outshine the host galaxy and show no evidence of spectral lines. The spectral energy distribution (SED) is composed of two main broad components \citep{1998MNRAS.299..433F}. The first one is extending from radio up to the X-rays, peaking in UV or soft X-rays. The second component extends up to multi-TeV energies, peaking in the hundreds of MeVs or GeVs. Depending on the frequency of the first component peak, blazars are subdivided into low-, intermediate- and high-frequency-peaked blazars (LBL, IBL, and HBL, respectively). Blazars show relentless variability in flux in all bands of the electromagnetic spectrum. Such variability time scales span from few minutes to months \citep{2015ApJ...811..143P}.

Mrk\,421 is a bright and close ($z = 0.031$) high-frequency-peaked blazar. It features bright and persistent $\gamma$-ray emission with frequent flaring activities. The SED of Mrk\,421 has been fitted with different emission models, e.g.: leptonic one-zone synchrotron self-Compton model (SSC) \citep{abdo_2011ApJ...736..131A}, hadronic model \citep{2015MNRAS.448..910C}. Observed fast variations hint at shocks moving relativistically within the jet or small size of emission region \citep{2014A&A...563A..91A} driven by the interactions in the jet or by magnetic reconnections. Numerous multi-wavelength (MWL) campaigns were performed to date observing Mrk\,421 from radio to the TeVs \citep[e.g.][]{abdo_2011ApJ...736..131A,2021A&A...647A..88A}. Mrk\,421 shows the highest variability in the X-rays and TeVs \citep{2021A&A...647A..88A}. The emission the X-ray and TeV band is highly correlated with close to zero days lag \citep{2021A&A...647A..88A, 2021MNRAS.504.1427A}.

Together with Mrk\,421, Mrk\,501 is one of the most prominent and well studies members of the blazars class. Mrk\,501 is also a bright and close ($z = 0.034$) source, showing strong and rapid variability down to few-minutes time scales \citep{2007ApJ...669..862A}. Based on its SED, Mrk\,501 is also classified as HBL showing a tendency to become extreme-high-frequency-peaked blazar (EHBL) during some flares \citep{2018A&A...620A.181A}. Mrk\,501 was the target of many multi-wavelength campaigns \citep[e.g.][]{2012ApJ...758....2B, 2018A&A...620A.181A}. Its SED can reasonably be explained by the one-zone SSC model \citep{2012ApJ...758....2B,2018A&A...620A.181A}, thought the introduction of a second smaller zone may be needed to explain features in the X-rays and TeV energies \citep{2018A&A...620A.181A}. The highest variability of Mrk\,501 appears to be in the TeV energies \citep{2018A&A...620A.181A}. Such behavior persists even if the periods of high activity are removed from the consideration.

In this paper, we report results from the ongoing multi-wavelength campaign. FACT monitoring for both sources already includes over eight years of data \citep{2013arXiv1311.0478D}. For this analysis, we limit the period to about five years, from December 2012 to April 2018. TeV observations were performed in unbiased (non-triggered) mode, and as regularly as possible given the observing conditions and technical constraints. Quasi-simultaneous data in radio, optical, UV, X-ray, and GeV bands were added to augment TeV observations to better characterize the multi-wavelength emission. The paper is structured as follows. In Sect.~\ref{sec:data}, an overview of the multi-instrument campaign is provided. The analysis of the variability in individual bands and cross-correlation behavior is presented in Sect~\ref{sec:timing}. In Sect~\ref{sec:conclusions}, we summarise and discuss the results.

\section{Multi-instrument data set\label{sec:data}}

The 5-year long multi-wavelength data-set consists of light curves from eight different instruments observing across the electromagnetic spectrum from radio up to TeV $\gamma$-rays. Light curves span from December 2012 to April 2018. To characterize the source variability and relation between the bands, the cross-correlation, Bayesian Block, and fractional variability analyses were performed. Both blazars were observed during various activity states and experience numerous flaring periods. One-day sampling in the TeV energies and sub-day sampling in the X-rays allowed to constrain the TeV-X-rays lag and identify individual flares. 

Observations in the TeV $\gamma$-rays were performed by the First G-APD Cherenkov Telescope (FACT). The telescope is located at Roque de los Muchachos Observatory (ORM) site operated by the Instituto de Astrofisica de Canarias (IAC) at La Palma \citep{2013JInst...8P6008A}. FACT is the first Imaging Air Cherenkov Telescope (IACT) adopting the silicon photo-multipliers (SiPMs) camera instead of vacuum-tube photo-multipliers in regular operation. The camera is comprised of 1440 pixels sampling at 2\,GHz. During normal data taking 300 samples (150\,ns) for each event are read out. The main segmented mirror has a reflective surface area of 9.5 m$^2$ (3.8\,m in diameter), providing a field-of-view of $4.5^{\circ}$. The telescope is in operation since October 2011 and is regularly taking the data. Since the commissioning, the control software was gradually automatized, resulting in remote operations since July 2012, and robotic since 2017. Such an operation scheme allows performing long-term regular observations of bright TeV sources with high cadence. A camera with SiPM and state-of-the-art feedback system in the camera allow operating in bright ambient conditions \citep{2013arXiv1311.0478D}. Description of the event reconstruction and data reduction technique is presented in \citep{2017ICRC...35..779H, 2021A&A...647A..88A,2017ICRC...35..608D}. The quality checks and background suppression techniques are described in \citep{2017ICRC...35..612M,2019arXiv190203875B,2019ICRC...36..630B}. The energy threshold based on simulated data is determined to be $\sim700$\,GeV and $\sim750$\,GeV for Mrk\,421 and Mrk\,501, respectively \citep{2021A&A...647A..88A}.

The LAT instrument onboard the Fermi Gamma-ray Space Telescope ({\it Fermi}-LAT) is the most sensitive $\gamma$-ray telescope in the 100\,MeV $<$ E $<$ 300\,GeV energy range. {\it Fermi}-LAT is using a charged particle tracker and a calorimeter to detect photons. The point spread function (PSF) depends on energy, reaching $1\sigma$-equivalent containment radius of $\sim 0.1^{\circ}$ at 40\,GeV \citep{2009ApJ...697.1071A}. Data reduction for both sources was performed using the PASS8 pipeline and Fermi Science Tool v10r0p5 package. Sources from {\it Fermi}-LAT 4-year Point Source Catalogue was used for the fitting model.

X-ray observations were performed in 15-50\,keV and 0.3-10\,keV bands independently by the Neil Gehrels Swift Observatory {\it Swift}/BAT and {\it Swift}/XRT instruments respectively. The reduction pipeline for the {\it Swift}/BAT data analysis is based on the BAT analysis software \texttt{HEASOFT} version 6.13 \cite{2013ApJS..207...19B}. The daily light curve for Mrk\,421 and three-day light curve for Mrk\,501 were built from 29344 orbital periods. {\it Swift}/XRT X-ray telescope \citep{2005SSRv..120..165B} light curves for Mrk\,421 and Mrk\,501 were downloaded from the official online Swift-XRT products generation tool\footnote{http://www.swift.ac.uk/user\_objects/}, which at the moment of the analysis was using the \texttt{HEASOFT} software version 6.22. Taking into account the high sensitivity of {\it Swift}/XRT, the light curves were separately built for two bands: 0.3-2\,keV and 2-10\,keV.

The {\it Swift} satellite also caries an ultra-violet optical telescope {\it Swift}/UVOT. The instrument is sensitive to light with a wavelength in 170-650 nm range. Using mechanically actuated filters, the telescope can observe in the following bands V, B, U, UVW1, UVM2, UVW2, and white (no-filter) \citep{roming_2005SSRv..120...95R}. For the observations reported in this paper, we consider data obtained only with UVW1, UVM2, and UVW2 filters. Since there is an overlap of the transparency windows of these three filters, we combine all the data points to enhance the source temporal coverage. The data reduction was performed using the on-/off-method with HEASOFT v6.24 software package with UVOT CALDB version 20170922. Both considered regions for the analysis were carefully checked in order not to include light from any nearby stars or UVOT mirror supporting structures. The dereddening of the light curves was performed assuming the reddening factors $E(B - V) = 0.013$ and $0.0164$ respectively for Mrk\,421 and Mrk\,501 \citep{2009ApJ...690..163R, 2011ApJ...737..103S}.

Optical observations in the V-band were performed by the 1.54\,m Kuiper Telescope on Mountain Bigelow and the 2.3\,m Bok Telescope on Kitt Peak as part of {\it Fermi} blazars monitoring campaign \citep{2009arXiv0912.3621S}. To cover the December 2012 - April 2018 time period, data from Cycle 5 to 10 were considered. The complete data set is publicly available\footnote{http://james.as.arizona.edu/$\sim$psmith/Fermi/DATA/photdata.html}.

Radio observations of Mrk\,421 and Mrk\,501, as part of {\it Fermi} blazars monitoring campaign, in 15\,GHz band were performed by the 40\,m Telescope of the Owens Valley Radio Observatory (OVRO) \citep{Richards_2011ApJS..194...29R}. The light curves with twice-per-week cadence are publicly available from the OVRO archive page\footnote{http://www.astro.caltech.edu/ovroblazars/}.

\section{Light curves timing and correlation analysis}
\label{sec:timing}

\subsection{Multi-wavelength variability}

To characterise the variability of a light curve, the fractional variability can be calculated using the relation $F_{var} = \sqrt{({S^2 - \langle\sigma^2_{err}\rangle})/{\langle{x}\rangle^2}}$ \citep{Vaughan_2003MNRAS.345.1271V}, where $S$ is the standard deviation of a flux, $\langle\sigma^2_{err}\rangle$ is the mean squared error, and $\langle{x}^2\rangle$ is the average flux squared.

\begin{figure}[h!]
  \centering
      \includegraphics[width=0.75\columnwidth]{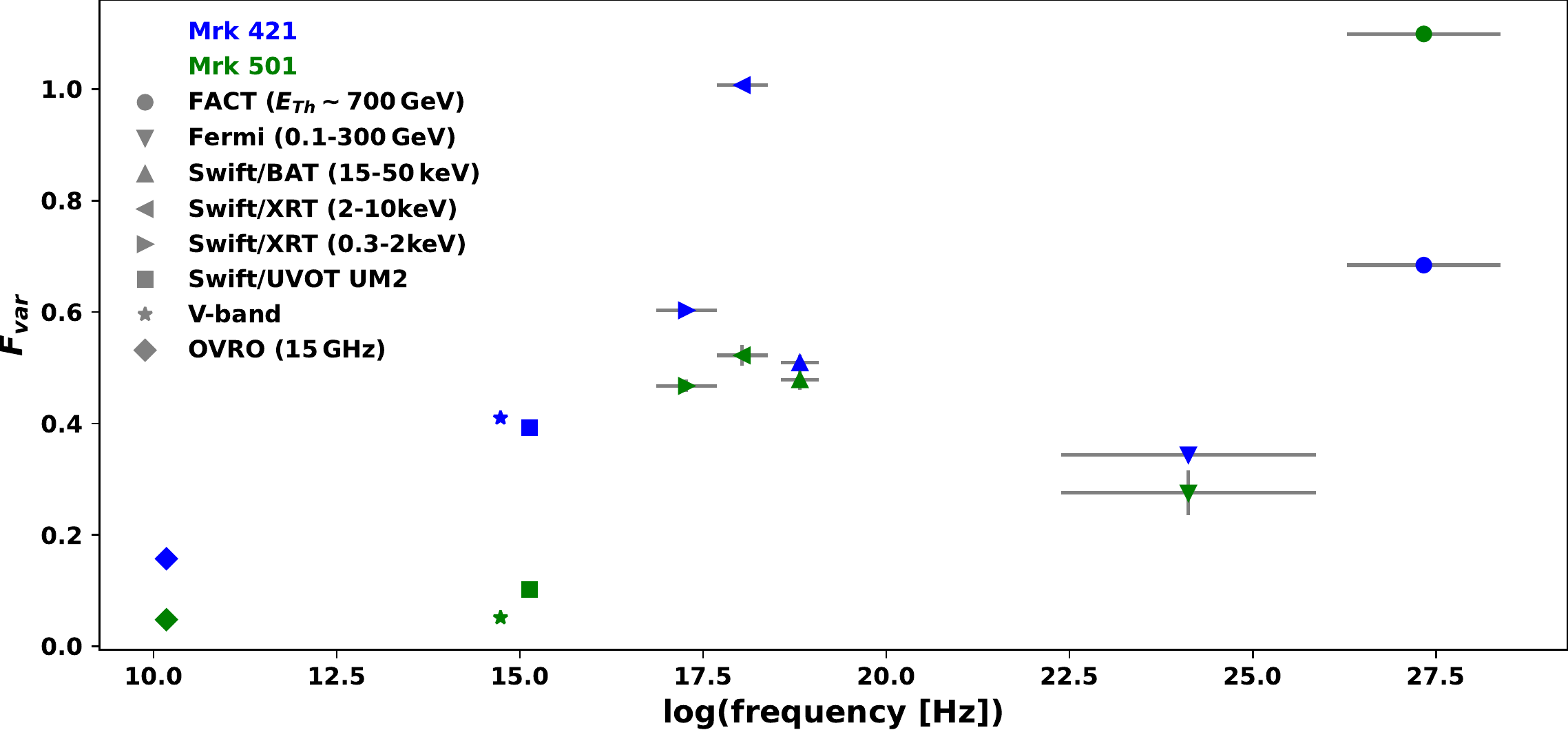}
  \caption{Fractional variability $F_{var}$ as a function of frequency for Mrk\,421 (blue markers) and Mrk\,501 (green markers). Horizontal and vertical lines denote the energy band and the $F_{var}$ uncertainties.}
  \label{fig:fvar}
\end{figure}

The uncertainties of $F_{var}$ are estimated following the prescription in \cite{Poutanen_2008MNRAS.389.1427P, Vaughan_2003MNRAS.345.1271V}. The $F_{var}$ value is generally sensitive to binning and data cleaning, reliability of $F_{var}$ is discussed in \citep{2015A&A...576A.126A,2019Galax...7...62S}. To exclude low-significance variations, we perform $F_{var}$ calculations for the light curves removing all the data points with significance below $2\sigma$.

Unlike previously reported monotonic increase of $F_{var}(\nu)$ dependency for Mrk\,501 \citep{2018A&A...620A.181A}, we found that for both sources (Fig.~\ref{fig:fvar}) fractional variability dependency on frequency has well pronounced two-hump structure, increasing from radio to X-rays, then dropping in {\it Fermi}-LAT band and increasing again in the TeV energies. Notably, the highest variability in Mrk\,421 is in 2-10\,keV X-ray band, while in Mrk\,501 it is in the TeVs reaching about the same value of $F_{var} \sim 1$. It is important to note that previous studies of the fractional variability in Mrk\,421 report three times lower values \citep{2015A&A...576A.126A, aleksic_2015A&A...578A..22A}, though such a discrepancy with the results presented in Fig~\ref{fig:fvar} may be explained by shorter light curves being used in those studies. The two-hump structure of $F_{var}(\nu)$ dependency indicates that the high-energy fraction of the two emission components is more variable than the low-energy ones. This suggests that the cutoff energies of both electron energy distributions (EED) are the primary source of the variability of the SED components.
 
\subsection{TeV -- X-ray correlations}

We use discrete correlation function (DCF)\citep{Peterson_1998PASP..110..660P} to investigate the connection between the emission in different bands and a lag between variations in those. DCF algorithm native uncertainties were used. Generally, in the case of short and noisy data-sets, such uncertainties are underestimated and may require advanced methods to understand the significance of the correlation \citep{2002MNRAS.332..231U}, albeit we use long-term light curves with a high signal-to-noise ratio (we also remove low significance ($<2\sigma$) data-points), so such consideration is not needed. 

\begin{figure}[h!]
  \centering
       \includegraphics[width=0.85\columnwidth]{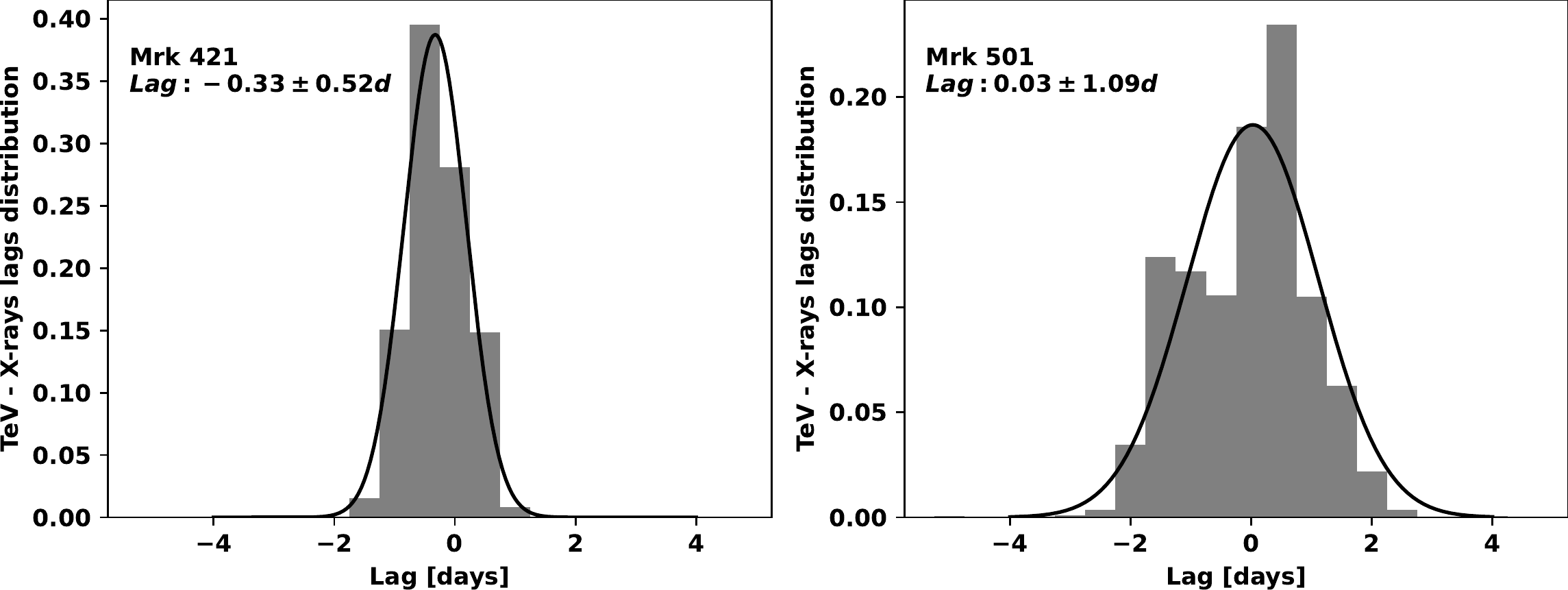}
       \caption{Combined lag distribution in Mrk\,421 (left) and Mrk\,501 (right) between TeV and X-ray ({\it Swift}/XRT 0.3-2\,keV and 2-10\,keV bands) light curves derived from FR/RSS DCF simulations. A Gaussian fit (black line) was applied to derive mean and uncertainty values.}
  \label{fig:dcf_all}
\end{figure}

The TeV and X-ray (Swift/XRT 0.3-2\,keV and 2-10\,keV bands) light curves (Fig. \ref{fig:dcf_all}) are strongly correlated, with close to zero lag. To estimate the lag, free of observational bias, we generated $10^4$ subsets for each pair of light curves using flux randomization (FR) and random subset selection (RSS) processes. For each such set of light curves, the DCF was calculated and the lag was recorded to built a lag distribution using a centroid threshold of 80\% of DCF maximum \citep{2004ApJ...613..682P}. The lag uncertainty corresponds to the standard deviation of the distribution of the lag values obtained for the random subsets. The cumulative lag derived from merging two lag distributions for both bands of {\it Swift}/XRT correlated with the TeV light curve is ($-0.33\pm0.52$) days and ($0.03\pm1.09$) days ($1\sigma$) for Mrk\,421 and Mrk\,501 respectively. For both blazars, the auto-correlations hint that the variability time scales are close to the light curve binning (1 day). Such variability and close to zero lag are consistent with predictions of one-zone SSC scenario, where X-rays are emitted during the synchrotron cooling of relativistic electrons, while the TeV photons are a result of inverse Compton scattering of photons on the same population of the electrons.

\subsection{Flares}

To identify individual flares in X-rays and TeV bands, the Bayesian Block algorithm \citep{2013ApJ...764..167S} was used. The algorithm was applied to TeV and X-ray light curves after $2\sigma$ cleaning was applied. Each identified block was considered to be a flare, if the average flux within a block is at least $2\sigma$ above the surrounding blocks and the duration of the block is at least two days, to ensure that at least two points are present in each block. The false-positive probability was set to 1\% \citep{2013ApJ...764..167S}. Neighboring blocks were merged, if both of them were above the threshold in respect to the further blocks on the left or/and on the right. In Mrk\,421, we identified 23 individual TeV flares, all of them, taking into account the coverage in X-rays (i.e. discarding the flare, if there are no data in the X-rays), were found to be coincident with the flares in the X-rays. In Mrk\,501, we identified 34 flares, and again all are coincident with the flares in the X-rays. Such a behavior, taking into account zero-compatible lag between the two bands, is expected within the SSC model.

\begin{figure}[h!]
  \centering
       \includegraphics[width=0.85\columnwidth]{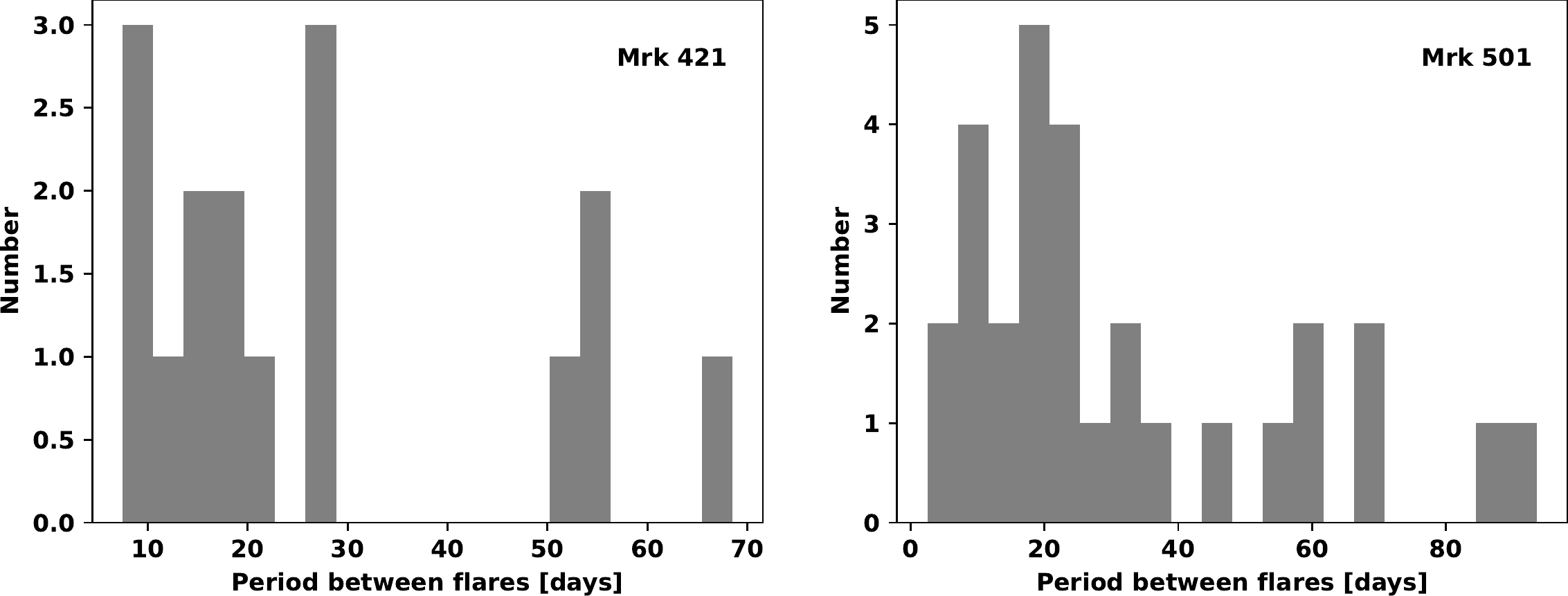}
       \caption{Time-interval between the TeV flares in Mrk\,421 (left) and Mrk\,501 (right).}
  \label{fig:flares_intervals}
\end{figure}

The time interval $\Delta t$ between FACT-detected flares shows a broad distribution for both blazars. In Mrk\,421, the most prominent peak of the distribution peaks between 7.5 days and 30 days (Fig.~\ref{fig:flares_intervals}, left). In Mrk\,501, the peak is wider spanning from 5 days to 40 days (Fig.~\ref{fig:flares_intervals}, right). The periods between the flares with a duration of over 60 days could be split in half, if a weak or missed flare was not detected between the two. Such periods between the flares may hint at the process driving the variability. Taking into account masses of SMBHs for both blazars, we find that the Lense–Thirring precession \citep{10.1093/mnrasl/slx174} for an inclined accretion disk \citep{1975ApJ...195L..65B} may be responsible for inducing the shocks which then travel downstream the jet.

\section{Conclusions\label{sec:conclusions}}

The analysis of 5-years long multi-wavelength light curves of Mrk\,421 and Mrk\,501 gives an interesting insight into the emission processes of these blazars. Most of the measured characteristics hint at the same processes being responsible for the multi-wavelength emission in the blazars. Both sources during the considered periods were found in a mixture of states, experiencing many flaring periods diluted with a quiescent state.

The fractional variability (highest variability in X-rays and TeV energies) of Mrk\,421 and Mrk\,501 along with the correlated TeV and X-ray emission indicate that the main source of variability is dominated by synchronous change of the cutoff energies of the low and high-energy components. A zero-compatible lag derived from correlations between the TeV and X-ray light curves of Mrk\,421 and Mrk\,501 indicates that emissions in these two bands are driven by the same physical parameter and are consistent with the leptonic emission scenario. This can be driven by variations of the electron maximal energy, or by e.g. the magnetic field that would affect both electrons and protons.

The observed periods between the TeV flares, 7.5-30 days in Mrk\,421, and 5-40 days in Mrk\,501, assuming a typical Doppler factor of $\delta=10$ and SMBH masses of $10^{8.5\pm 0.18}M_{\odot}$ and $10^{9.2 \pm 0.3}M_{\odot}$, respectively, are compatible with predictions of the Lense–Thirring precession for an inclined accretion disk.

\noindent\textit{Acknowledgements.} This research has made use of public data from the \textit{OVRO} 40-m telescope \citep{Richards_2011ApJS..194...29R}, the Bok Telescope on Kitt Peak and the 1.54 m Kuiper Telescope on Mt. Bigelow \citep{2009arXiv0912.3621S}, \textit{Fermi}-LAT \citep{2009ApJ...697.1071A} and  \textit{Swift} \citep{2004NewAR..48..431G}. Acknowledgements of the FACT Collaboration are availabe at \href{https://fact-project.org/collaboration/icrc2021_acknowledgements.html}{https://fact-project.org/collaboration/icrc2021\_acknowledgements.html}.

\vspace*{-0.7em}
\bibliographystyle{JHEP}
\setlength{\bibsep}{0.3pt}
\footnotesize{
\bibliography{10_references}
}
\end{document}